\documentclass[a4paper,12pt]{article}

\usepackage[T1]{fontenc}
\usepackage[latin2]{inputenc}
\usepackage{graphicx}
\input{epsfx}
\usepackage{subfig}
\usepackage{wrapfig}
\usepackage{amsmath}
\usepackage{amsfonts}
\usepackage{amssymb}
\usepackage{color}
\usepackage{rotating}
\usepackage{multirow}

\newcommand{\be}{\begin{equation}}
\newcommand{\ee}{\end{equation}}
\newcommand{\bes}{\begin{subequations}}
\newcommand{\ees}{\end{subequations}}
\newcommand{\bea}{\begin{eqnarray}}
\newcommand{\eea}{\end{eqnarray}}
\newcommand{\bear}{\begin{equation}\begin{array}}
\newcommand{\eear}[1]{\end{array}\label{#1}\end{equation}}
\def\ba{$$\begin{array}}
\def\ea{\end{array}$$}
\def\bra{$\begin{array}}
 \def\era{\end{array}$}

\definecolor{Red}{rgb}{1,0,0}
\definecolor{Blue}{rgb}{0,0,1}

\newcommand{\fr}[2]{\dfrac{{ #1}}{{ #2}}}

\newcommand{\la}{\langle}
\newcommand{\ra}{\rangle}
\newcommand{\fn}[1]{\footnote{{\sf #1}}}

\newsavebox{\fmbox}

{\end{list}}
\newcounter{enumct}

\begin{document}
\renewcommand{\tilde}{\widetilde}
\renewcommand{\ge}{\geqslant}
\renewcommand{\le}{\leqslant}

\title{Temperature evolution of physical parameters in the Inert Doublet Model}

\author{Dorota Soko\l owska \\
\textit{University of Warsaw, Faculty of Physics, Warsaw, Poland}
 }


\maketitle

\begin{abstract}

Inert Doublet Model is a minimal extension of the Standard Model with the second scalar doublet that may provide a Dark Matter candidate. In this paper we consider different variants of the evolution of the Universe after inflation, that lead towards the Inert phase today. We extend our previous analysis, in particular by discussing the co-existence of minima and providing numerical examples of the evolutions of the mass parameters for the different types of the history of the Universe. We take into account the existing constraints, including the relict density data.

\end{abstract}

\section{Introduction}

The Inert Doublet Model (IDM) \cite{inert} is a $Z_2$ symmetric 2HDM, which for a special set of parameters may provide the Dark Matter (DM) candidate.  The model contains two scalar $SU(2)$ doublets.  One is a "standard" scalar (Higgs) doublet $\Phi_S$, responsible for electroweak symmetry breaking and masses of fermions and gauge bosons as in the Standard Model (SM), with a Higgs particle $h_S$.
The second one is a "dark" scalar doublet $\Phi_D$, which does not receive  vacuum expectation value (v.e.v.) and does not couple to fermions\fn{Our notations are similar to those in the general 2HDM with the change $\Phi_1\to\Phi_S$, $\Phi_2\to\Phi_D$.}. In the model the following discrete symmetry of the $Z_2$ type, which we call \textit{the $D$-symmetry}, is present\fn{$SM$ denotes SM fermions $\psi_f$ and gauge bosons.}:
\begin{equation}
D: \quad \Phi_S \xrightarrow{D} \Phi_S,\quad
	\Phi_D \xrightarrow{D} -\Phi_D,\quad
	SM     \xrightarrow{D} SM.
	\label{dtransf}
\end{equation}

All the components of the dark doublet $\Phi_D$ are realized
as the massive  $D$-scalars: two charged $D^\pm$ and two neutral
ones $D_H$ and $D_A$. By construction, they possess a conserved multiplicative quantum number, \textit{the $D$-parity},
and therefore the lightest particle among them can be considered as a candidate for the DM particle.

In this paper we discuss the evolution of the Universe during its cooling down  after inflation, following the approach presented in \cite{GK07, GIK09, iv2008, nasza}. As before, we assume that the current state of the Universe is described by IDM. In this analysis we include all existing constraints on the model, together with the corresponding energy relict density, calculated by us \cite{Omega-DS} using micrOMEGAs \cite{micro}.

In sec. 2 we list the basic properties of IDM and discuss the relevant astrophysical and collider constraints. Sec. 3 contains the summary of the basic assumptions of our approach and extended analysis of the possible today's states and types of evolution, as compared to \cite{nasza}. In particular we trace the co-existence of minima. In sec. 4 we provide numerical examples of temperature evolution of physical parameters, like masses. 

\section{IDM}\label{idm}

\subsection{Lagrangian}
We consider
an electroweak symmetry breaking (EWSB) via the Brout-Englert-Higgs-Kibble (BEHK) mechanism described by the Lagrangian
\begin{equation}
{ \cal L}={ \cal L}^{SM}_{ gf } +{ \cal L}_H + {\cal L}_Y(\psi_f,\Phi_S) \,, \quad { \cal L}_H=T-V\, .
\label{lagrbas}
\end{equation}
Here, ${\cal L}^{SM}_{gf}$ describes the $SU(2)\times U(1)$ Standard Model
interaction of gauge bosons and fermions, which is independent on the
realization of the BEHK mechanism.
In the considered case the scalar Lagrangian ${\cal L}_H$ contains the standard kinetic term $T$
and the $D$-symmetric potential $V$, which can describe IDM, with two scalar  doublets $\Phi_S$ and $\Phi_D$:
\begin{eqnarray}
V = -\frac{1}{2}\left[m_{11}^2 \Phi_S^\dagger\Phi_S + m_{22}^2 \Phi_D^\dagger\Phi_D \right] + \frac{1}{2}\left[\lambda_1 \left(\Phi_S^\dagger\Phi_S\right)^2 + \lambda_2 \left(\Phi_D^\dagger\Phi_D \right)^2\right] \nonumber \\
+  \lambda_3 \left(\Phi_S^\dagger\Phi_S \right) \left(\Phi_D^\dagger\Phi_D\right) + \lambda_4 \left(\Phi_S^\dagger\Phi_D\right) \left(\Phi_D^\dagger\Phi_S\right) +\frac{1}{2}\lambda_5\left[\left(\Phi_S^\dagger\Phi_D\right)^2\!+\!\left(\Phi_D^\dagger\Phi_S\right)^2\right]. \label{pot}
\end{eqnarray}
All parameters in $V$ are taken to be real, with $\lambda_5 <0$ \cite{nasza}. 
For the further discussion, it is useful to introduce the $(\mu_1,\mu_2)$ phase space with:
\begin{eqnarray}
 &\mu_1 = m_{11}^2/\sqrt{\lambda_1} , \quad  \mu_2= m_{22}^2/\sqrt{\lambda_2}.&
\end{eqnarray}

\textit{Positivity conditions} imposed on the potential guarantee the existence of the stable vacuum. They assure that the potential is bounded from below, meaning that the extremum with the lowest energy will be the global minimum of the potential (vacuum). The positivity constrains relevant for this analysis are:
\begin{eqnarray}
& \lambda_1>0\,, \quad \lambda_2>0, \quad R + 1 >0, & \label{posit}\\[1mm]
&  R = \lambda_{345}/\sqrt{\lambda_1 \lambda_2}, \quad \lambda_{345}=\lambda_3+\lambda_4+\lambda_5. \label{Rdef} &
\end{eqnarray}

 ${\cal L}_Y$ describes the Yukawa interaction of SM fermions $\psi_f$
with only one scalar doublet $\Phi_S$, having the same form as in the SM with the change $\Phi\to\Phi_S$ (Model I for a general 2HDM). For quarks it reads:
\begin{equation}
{\cal L}_Y = \bar{Q}_L \Gamma \Phi_S d_R + \bar{Q}_L \Delta \tilde{\Phi}_S u_R + (h.c),
\end{equation}
where $Q_L$ is doublet of left-handed quarks, $d_R,u_R$ are the right-handed quarks, $\Gamma$ and  $\Delta$ are $3\! \times\! 3$ matrices of Yukawa couplings in generations space, $\tilde \Phi=i \sigma_2 \Phi^*$.  Similar Yukawa term is introduced 
for leptons, namely $\bar{L}_L \Gamma \Phi_S l_R + (h.c)$.
${\cal L}_Y$ respects $D$-symmetry in any order of the perturbation theory.

\subsection{Inert vacuum state \label{sec:vacinert}}

Inert extremum, denoted by $I_1$, is realized if the extremum conditions for the potential $V$ respect

\begin{equation}\label{in_extr}
v^2 = \mu_1/ \sqrt{\lambda_1}.
\end{equation}

$I_1$ realizes vacuum if following conditions are satisfied \cite{nasza}:
\begin{eqnarray}
 \mu_{1} >0 \textrm{ for any } R, \quad \mu_1 > \mu_2 \textrm{ for } R>1, \quad R \mu_1 > \mu_2 \textrm{ for } |R|<1.
\end{eqnarray}

%
There are four dark scalar particles $D_H,\,D_A,
D^\pm$ and the Higgs particle $h_S$ which interacts
with the fermions and gauge bosons just as the Higgs boson in the SM.

Inert state  is invariant under the $D$-transformation just as the whole basic Lagrangian \eqref{lagrbas} does. Therefore, the  $D$-parity is conserved and  due to this fact
the lightest $D$-odd particle is stable, being a good  DM candidate.

Masses of the scalar particles are:
\bear{c}
M_{h_s}^2=\lambda_1v^2= m_{11}^2\,,\qquad M_{D^\pm}^2=\fr{\lambda_3 v^2-m_{22}^2}{2}\,,\\[3mm]
M_{D_A}^2=M_{D^\pm}^2+\fr{\lambda_4-\lambda_5}{2}v^2\,,\qquad M_{D_H}^2=
M_{D^\pm}^2+\fr{\lambda_4+\lambda_5}{2}v^2\,.
\eear{massesA}
Assuming, as usual, that DM particles are neutral,
we consider such  variant of IDM, in which
\be
M_{D^\pm}, \; M_{D_A} > M_{D_H}.
\label{chargedheavy}
\ee

After EWSB parameters of the potential $V$ can be expressed by the four scalar masses and two self-couplings $\lambda_{345}, \, \lambda_2$ between the neutral scalar  particles. \textit{Triple and quartic} couplings between SM-like Higgs $h_S$ and DM candidate $D_H$, i.e. $D_H D_H h_S$ and $D_H D_H h_S h_S$, are proportional to $\lambda_{345}$. The second coupling, $\lambda_2$, is related only to a \emph{quartic} self-coupling, $D_H D_H D_H D_H$. The remaining self-coupling, $\lambda_3$, governs the charged scalars' interactions: $D^+ D^- h_S$ and $D^+ D^- h_S h_S$.

\subsection{Constraints \label{sec:constr}}

Various  theoretical and experimental constraints apply for the IDM (see e.g.~\cite{limpap}-\cite{kra-sok}). 

The value of $\lambda_{345}$ strongly affects the DM interactions relevant for the DM energy relict density $\Omega_{DM} h^2$. In general, for larger $|\lambda_{345}|$ the relict density decreases due to the enhanced $D_H D_H$ annihilation via s-channel Higgs exchange. This parameter also plays an important role in the indirect detection of DM \cite{Dolle:2009fn,LopezHonorez:2006gr} and larger $|\lambda_{345}|$  gives an enhanced flux of neutrinos and gamma rays.

The value of the remaining coupling, $\lambda_2$, does not influence the DM relict density explicitly and so this parameter is usually fixed to arbitrary small value during the DM analysis of IDM \cite{Dolle:2009fn,LopezHonorez:2006gr}. However, value of $\lambda_2$ limits the value of $\lambda_{345}$ via the positivity constraints ($\lambda_2>|\lambda_{345}|/\sqrt{\lambda_1}$) and therefore it plays an important role in the analysis, as discussed in \cite{Omega-DS}.

%

\paragraph{Collider constraints on scalars' masses} Electroweak precision tests constrain strongly physics beyond SM. For IDM both light and heavy Higgs particle is allowed \cite{inert}. Constraints on the mass splittings  $\delta_A = M_{D_A} - M_{D_H}, \, \delta_{\pm} = M_{D^\pm} - M_{D_H}$ have been discussed in \cite{Dolle:2009fn}. For a light Higgs boson, the allowed region corresponds to  $\delta_{\pm} \sim \delta_{A} $ with mass splittings that could be large.
 For heavy SM Higgs large $\delta_{\pm}$ is needed, while $\delta_{A} $ could be small. In this work we limit ourselves to the light SM-like Higgs boson.

As $D^\pm, D_A$ and $D_H$ do not couple to fermions, the LEP limits based on Yukawa interaction for the standard 2HDM don't apply. However, the signatures are similar to neutralinos and charginos interactions in MSSM and the absence of a signal at LEP II was interpreted within the IDM  in paper \cite{Lundstrom:2008ai}. This analysis excludes the following region of masses: $M_{D_H} < 80$ GeV, $M_{D_A} < 100$ GeV and $\delta_A > 8$ GeV. For $\delta_A <8$ GeV the LEP I limit $M_{D_H} + M_{D_A} > M_Z$ applies.

\paragraph{Constraints on self-couplings}  
The positivity constraints are imposed directly on quartic parameters in the potential (\ref{posit},\ref{Rdef}). If we want to assure the perturbativity of the theory, self-couplings $\lambda$ cannot be large. The bound (called \textit{perturbativity constraint}) is set typically to
\begin{equation}
|\lambda|<4 \pi.
\end{equation}

Astrophysical estimations of the energy relict density $\Omega_{DM} h^2$ may be used to give the limitations for $|\lambda_{345}|$ depending on the chosen value of masses of $D_H$ and other scalars \cite{Dolle:2009fn,LopezHonorez:2006gr}. However, they do not directly constrain the remaining quartic coupling $\lambda_2$.

\paragraph{DM relict density constraints} 

The DM energy density in the Universe is estimated to \cite{PDG}:

\begin{equation}
\Omega_{DM}h^2=0.112 \pm 0.009.\label{DMdens}
\end{equation}

In this analysis we assume that $D_H$ is a dominant component of the observed DM and density \eqref{DMdens} is today's density of $D_H$. 

Various studies \cite{Dolle:2009fn,LopezHonorez:2006gr} and our independent analysis show that for IDM there are three allowed regions of $M_D$: (i) light DM particles with mass close to and below $10 \textrm{ GeV}$, (ii) medium mass regime of $40-80 \textrm{ GeV}$ and (iii) heavy DM of mass larger than $500 \textrm{ GeV}$. For purpose of this paper we concentrate on the medium mass region with the chosen set of masses (sec.4). With this choice parameters $\lambda_2$ and $\lambda_{345}$ are free (up to the limitations discussed above). We consider different variants of their choice corresponding to different types of the evolution of the Universe.

For our analysis use of standard available tools, ie. micrOMEGAs, is in fact  limited, as this program neglects temperature dependence of physical parameters and a possibility of more than one phase transition. In paper \cite{nasza} we concluded that if in the past there were sequences of phase transitions, then the Universe entered the inert phase with DM candidate at lower temperatures that in one-stage EWSB. This should be taken into account while solving the Boltzmann equations for DM relict density $n_D$. Furthermore, one should consider the latent heat of the first order transition or strong fluctuations in the sequences of the second order phase transitions \cite{pracaTEMP}. For a single phase transition the use of present form of micrOMEGAs  is more justified, however taking into account the evolution of masses and existence of different decay channels may provide corrections. In this sense, the relict density calculations in this paper should be considered as a preliminary estimate.

\section{Thermal evolution of the Universe}\label{evolution}

We consider thermal evolution of the Lagrangian,
following the approach presented in \cite{iv2008,GIK09,nasza}. It allows to study the earlier history of the Universe after inflation. In the   first nontrivial approximation the Yukawa couplings and  the quartic coefficients $\lambda's$  remain unchanged, while the mass parameters $m_{ii}^2 \; (i=1,2)$ vary with temperature $T$ as follows\fn{Formulae for $c_1,c_2$ were obtained recently by G. Gil (Master Thesis, 2011); they correct the corresponding formulae given in \cite{nasza}.}:
\bear{c}
m_{ii}^2(T)=  m_{ii}^2-c_iT^2,\\[3mm]
c_1=\fr{3\lambda_1+2\lambda_3+\lambda_4}{6}+\fr{3g^2+g^{\prime 2}}{8}+\fr{g_t^2+g_b^2}{2}, \quad
c_2=\fr{3\lambda_2+2\lambda_3+\lambda_4}{6}+\fr{3g^2+g^{\prime 2}}{8}.			 \eear{Tempdep}

Here $g$ and $g^\prime$ are  the EW gauge couplings, while $g_t$ and $g_b$ are the SM Yukawa
couplings  for $t$ and $b$ quarks, respectively\fn{Normalization of couplings: $g_i = \sqrt{2} m_i/v, (g_t \approx 0.99, g_b \approx 0.02)$; $g= 2 M_W /v = 0.652, g'=0.351$.}.

In virtue of positivity conditions the sum of evolution coefficients is positive: $c_2+c_1>0$. For $R>0$ both $c_i>0$, while for $R<0$ arbitrary signs of $c_{1,2}$ are possible. In this work we limit ourselves to positive $c_1,c_2$ as we consider only the restoration of EW symmetry for high $T$ (this corresponds to the negative values of $m_{11}^2(T), \; m_{22}^2(T)$ for high enough $T$) \cite{Gavela:1998ux}. See \cite{nasza} for more details.

\subsection{Possible minima during evolution}\label{extrema}

As the Universe is cooling down the potential $V$ (\ref{pot}), with temperature dependent quadratic coefficients (\ref{Tempdep}), may have different ground states. The general possible extrema are in form:
\bear{c}
        \langle\Phi_S\rangle =\dfrac{1}{\sqrt{2}}\left(\begin{array}{c} 0\\
        v_S\end{array}\right),\quad \langle\Phi_D\rangle
        =\dfrac{1}{\sqrt{2}}\left(\begin{array}{c} u \\ v_D
        \end{array}\right),
\eear{genvac}
with $v_S >0$ and $v^2=v_S^2+|v_D^2|+u^2$.


 \begin{table}
{\renewcommand{\arraystretch}{2}

\begin{tabular}{|c|p{7.5cm}|p{4.5cm}|}
\hline
\hline
\multicolumn{3}{|c|}{Extrema and vacua} \\ \hline \hline
name of extremum &properties of vacuum & vev's \\\hline
\textit{EW symmetric}: $EW\! s$  & Massless fermions and bosons and massive scalar doublets. & $v_D=0, \quad v_S=0$ \\ \hline
\multirow{2}{*}{ \textit{inert}: $I_1$}  & \multirow{2}{7.7cm}{Massive fermions and gauge bosons; scalar sector: SM-like Higgs $h_S$ and dark scalars $D_H, D_A, D^\pm$ with \textbf{DM candidate} $D_H$.} & $v_D=0$,\\ & & $v_S^2=v^2=\fr{\mu_{1}}{\sqrt{\lambda_1}}$\\ \hline
\multirow{2}{*}{ \textit{inertlike}: $I_2$}  & \multirow{2}{7.7cm}{Massless fermions and massive gauge bosons; scalar sector: Higgs particle $h_D$ (no interaction with fermions), four scalars $S_H,S_A,S^\pm$, \textbf{no DM candidate}.} & $v_S=0$,\\ & & $v_D^2=v^2=\fr{\mu_{2}}{\sqrt{\lambda_2}}$ \\ \hline
\multirow{3}{*}{\textit{mixed}: $M$} & \multirow{3}{7.7cm}{Massive fermions and bosons, 5 Higgs particles: CP-even $h$ and $H$, CP-odd $A$ and charged $H^\pm$, \textbf{no DM candidate}.} & $v^2=v_S^2+v_D^2$,\\ & & $v_S^2=\fr{\mu_1-R\mu_2}{\sqrt{\lambda_1}(1-R^2)} >0$,\\
& & $v_D^2=\fr{\mu_2-R\mu_1}{\sqrt{\lambda_2}(1-R^2)} >0$\\ \hline
\hline
\end{tabular}
}
  \caption{General properties of the extrema and vacua, following \cite{nasza}. \label{vacua}}
\end{table}

Possible neutral solutions ($u=0$) are: EW symmetric $EW\! s$, inert $I_1$, inertlike $I_2$ and mixed $M$. Their general properties are summarized in table~\ref{vacua} \cite{nasza}, see also the appendix A. 

There are three distinguish allowed regions of parameters $\lambda's$, the best parametrized by the parameter $R$ (\ref{Rdef}), namely  a) $R>1, \;$b) $ 1>R>0, \;$c) $0>R>-1$. 

The EW symmetric state ($EW\! s$) exists for every value of $R$ if and only if both $\mu_1< 0$ and $\mu_2< 0$, being the only existing extremum (and thus the vacuum).
For $R>1$ (fig.\ref{wykresa}) the energy of the mixed extremum $M$ (if it exists) is always higher than for the other extrema, so it cannot be the vacuum \cite{nasza}. Possible $EW\! v$ vacua in this case are $I_1$ and $I_2$.
Fig.\ref{wykresb} shows the allowed regions for $0<R<1$. Again we have the regions of $EW\! s$, $I_1$ and $I_2$ vacua, but now also mixed extremum $M$ can be realized as a vacuum.
Case of $R <0$ is similar to the previous one but with the wider region of $M$ vacuum (fig.\ref{wykresc}). Summary of existing extrema, local minima and vacua for various $R$ regions can be found in table 2.

\begin{table}[h]
{\renewcommand{\arraystretch}{1.5}

\begin{tabular}{|c|p{12cm}|}
\hline
\hline
\multicolumn{2}{|c|}{\textbf{a) Extrema and  vacua for ${\pmb R} \, {\pmb >} \, {\pmb 1}$}} \\ \hline \hline
$\mu_2 > 0 \textrm{ and } \mu_2 > \mu_1 R$ & $I_2$ is the vacuum. For $\mu_1>0$  $I_1$ is an extremum, but not a minimum. \\ \hline
 $0< \mu_1 < \mu_2 < \mu_1R$ & $I_2$ is the vacuum (i.e. global minimum) and $I_1$ -- a local minimum.
\\ \hline
$0< \mu_1 R^{-1} < \mu_2 < \mu_1$ & $I_1$ is the vacuum (i.e. global minimum) and $I_2$ -- a local minimum.\\ \hline
$\mu_1 > 0 \textrm{ and } \mu_2 < \mu_1 R^{-1}$ & $I_1$ is the vacuum. For $\mu_2 >0$ $I_2$ is an extremum, but not a minimum.
 \\ \hline
\hline

\hline
\multicolumn{2}{|c|}{\textbf{b) Extrema and  vacua for $ {\pmb  1} \, {\pmb >} \, {\pmb R} \, {\pmb >} \, {\pmb 0} $}} \\ \hline \hline
 $\mu_2 > 0 \textrm{ and } \mu_2 > \mu_1 R^{-1} $ &  $I_2$ is the vacuum. For $\mu_1>0$ $I_1$ is an extremum, but not a minimum.\\ \hline
 $\mu_1> 0 \textrm{ and } \mu_2 < \mu_1 R $ &  $I_1$ is the vacuum. For $\mu_2>0$ $I_2$ is an extremum, but not a minimum.
 \\ \hline
$0< \mu_1 R < \mu_2 < \mu_1 R^{-1}$ & $M$ is the vacuum, $I_1$ and $I_2$ are the extrema.\\ \hline
\hline

\hline
\multicolumn{2}{|c|}{\textbf{c) Extrema and  vacua for ${\pmb 0 } \, {\pmb >} \, {\pmb R} \, {\pmb >} \, {\pmb -} {\pmb 1} $}} \\ \hline \hline
 $\mu_2> 0 \textrm{ and } \mu_2< \mu_1 R^{-1}$ &
$  I_2$ is the vacuum, no other extrema. \\ \hline
 $\mu_1> 0 \textrm{ and } \mu_2 < \mu_1 R $  & $I_1$ is the vacuum, no other extrema. \\ \hline
$\mu_2 > \textrm{Max} \left( \mu_1 R^{-1}, \mu_1 R \right) $ &
$M$ is the vacuum. Extrema: $I_2$ for $\mu_2>0$, $I_1$ for $\mu_1>0$. \\ \hline
\hline
\end{tabular}
}
  \caption{$EW\! v$ extrema and  vacua for various $R$ regions. \label{vacuaR}}
\end{table}

\subsection{Possible sequences of phase transitions }\label{sequences}

We use ($\mu_1(T),\mu_2(T)$) phase diagrams and the redefined evolution coefficients $\tilde c_1, \, \tilde c_2$ to determine the sequences of transitions between different vacua as $T$ decreases:
\begin{eqnarray}
 &\mu_1(T) = m_{11}^2(T)/\sqrt{\lambda_1} , \quad  \mu_2(T) = m_{22}^2(T)/\sqrt{\lambda_2},&\\[2mm]
&\tilde c_1=c_1/\sqrt{\lambda_1},
\qquad \tilde c_2=c_2/\sqrt{\lambda_2}, \qquad \tilde c=\tilde c_2/\tilde c_1.&\label{ctilde}
\end{eqnarray}

\begin{figure}[htb]
\vspace{-10pt}
  \centering
  \subfloat[$R>1$]{\label{wykresa}\includegraphics[width=0.3\textwidth]{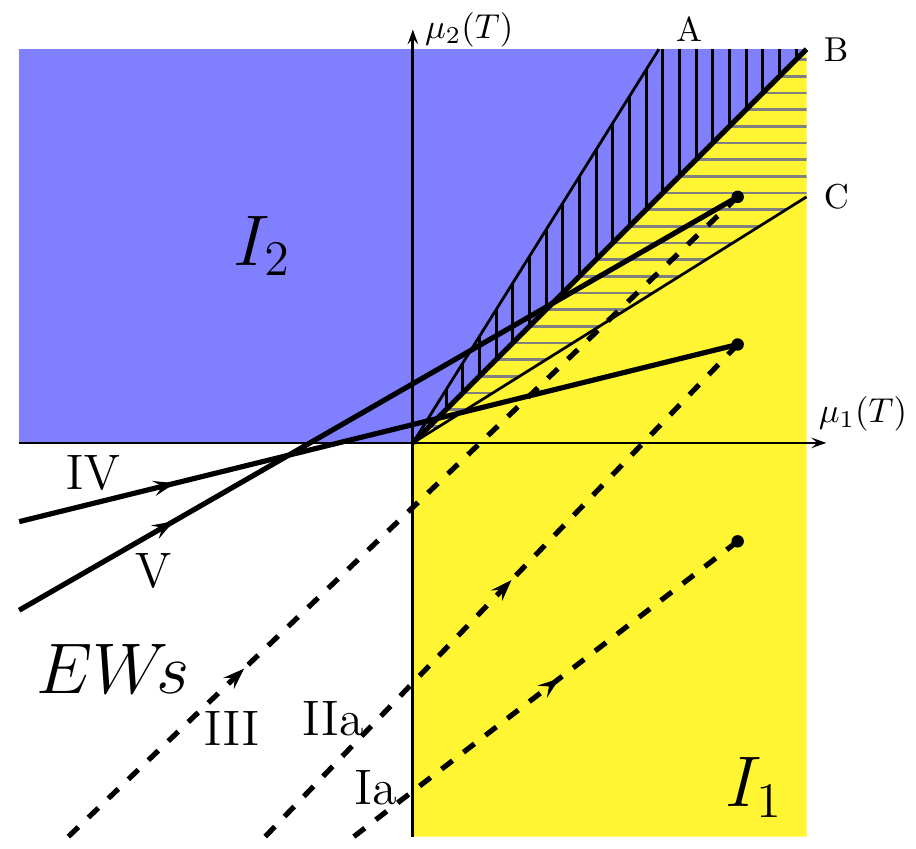}} \quad
  \subfloat[$1>R>0$]{\label{wykresb}\includegraphics[width=0.3\textwidth]{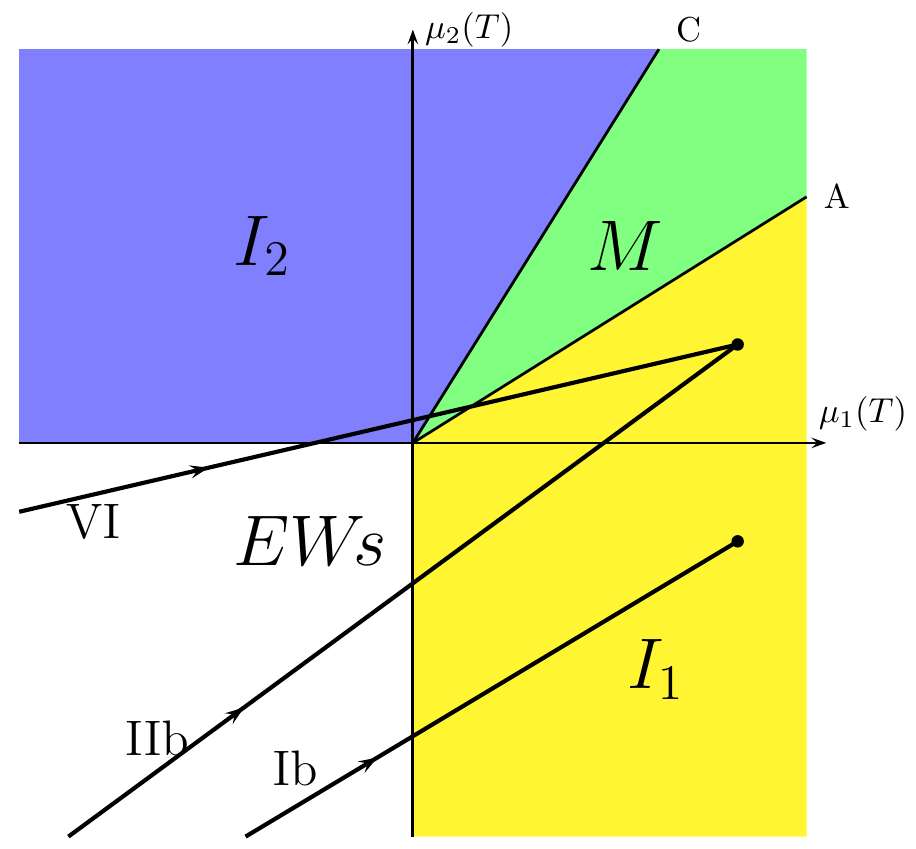}} \quad
  \subfloat[$0>R>-1$]{\label{wykresc}\includegraphics[width=0.3\textwidth]{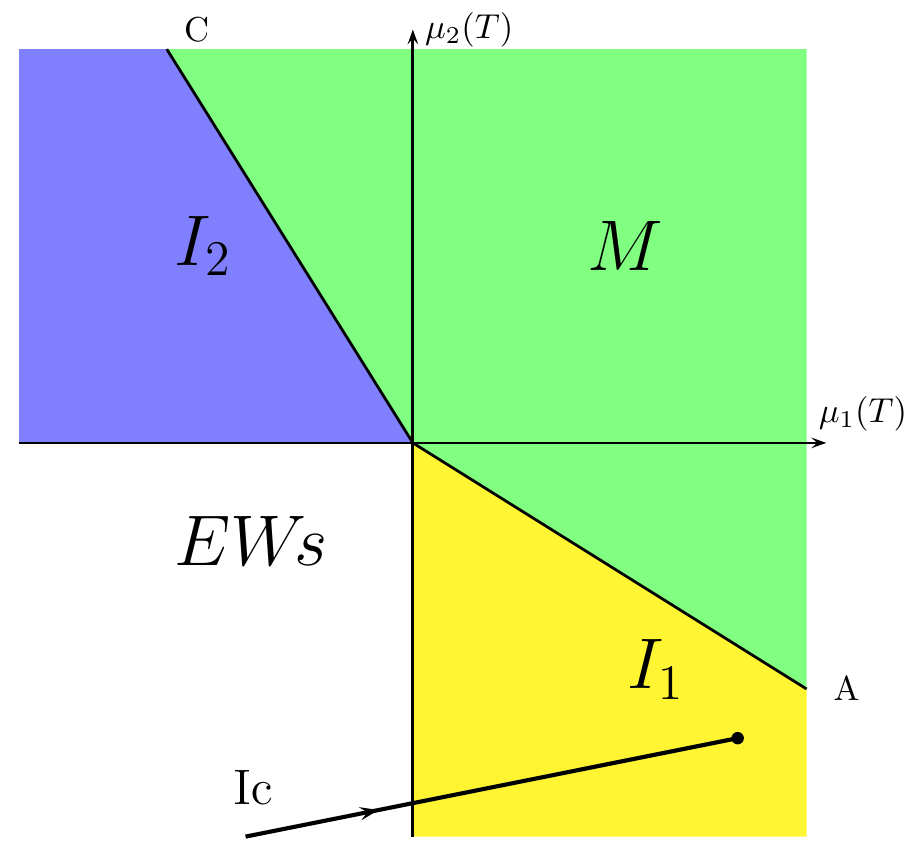}}
\vspace{-5pt}
  \caption{  Possible vacua and evolution to a current states (dots) represended by rays on $(\mu_1,\mu_2)$ plane for a) $R>1$, b) $1>R>0$, c) $
0>R>-1$. The boundary lines are $A:\mu_2 = \mu_1 R$,\, $B:\mu_2 = \mu_1$,
$\, C:\mu_2 = \mu_1/ R$. Blue (dark shade) region represents $I_2$ vacuum,
yellow (light shade) region -- $I_1$ vacuum and green (medium shade)
region -- $M$ vacuum. In the hatched regions between lines $A,B$ and $B,C$ $I_1$ and $I_2$
minima co-exist. 
   }
  \label{fig:rays}
\end{figure}

Figs.(\ref{fig:rays}a,b,c) show all possible types of evolution from the EW symmetric phase towards the inert phase today. The possible evolutions are represented by the rays directed towards  the growth of time, ie. the decrease of the temperature. Today's values are defined by $\mu_1=\mu_1(0),\mu_2=\mu_2(0)$ and marked by dots in figures.

The relevant temperatures with the corresponding conditions for the phase transitions are \cite{nasza}:

\be
 T_{EW\! s,1}=\sqrt{\mu_1/\tilde c_1} \qquad (\mu_1(T_{EW\! s,1})=0), \label{TEWSBI_1}
\ee
\be
T_{EW\!s,2}=\sqrt{\mu_2/\tilde c_2} \qquad (\mu_2(T_{EW\! s,2})=0), \label{TEWSBI_2}
\ee
\be
T_{2,1}=\sqrt{ \fr{\mu_1-\mu_2}{\tilde c_1 -\tilde c_2}} \qquad (\mu_1(T_{2,1})=\mu_2(T_{2,1})), \label{TI1I2}
\ee

\bear{c}
   T_{2,M}= \sqrt{\fr{\mu_1-R\mu_2}{\tilde c_1 - R\tilde c_2}}, \quad  T_{M,1}= \sqrt{\fr{R\mu_1-\mu_2}{R\tilde c_1 - \tilde c_2}}\,,\\[4mm]
   \left(\mu_2(T_{2,M})=\mu_1(T_{2,M})/R, \quad \mu_1(T_{M,1})=\mu_2(T_{M,1})/R\right)\,.
\eear{tempIN}

\subsubsection{$R>1$}

For $R>1$ the possible types of evolution that lead towards the inert phase today can be divided into two groups (fig.\ref{wykresa}, table~\ref{tabler1}). The first one corresponds to a single phase transition of the 2nd-order at the temperature (\ref{TEWSBI_1}), shown by  rays Ia, IIa and III.
The second type of evolution is in form of a sequence of two phase transitions represented by the  rays IV and V. For these rays, the first phase transition $EW\! s \to I_2$ (EWSB) realized at temperature (\ref{TEWSBI_2}) is a 2nd-order phase transition, while the second one $I_2 \to I_1$, realized at temperature (\ref{TI1I2}), is a 1st-order transition.

We notice that the considered $R>1$ case contains an unique opportunities of co-existing minima (vacuum $I_1$ and local metastable minimum $I_2$) for rays III, IV and V. Also, only in this case there is a possibility of the 1st-order phase transition (rays IV and V). That cannot be realized for the other values of $R$. For ray IV the co-existence is temporary and the local minimum $I_2$ disappear for low temperatures, while for rays III and V it still exists for $T=0$.

For $R=1$ the phase space of co-existing minima $I_1$ and $I_2$ is reduced to line $\mu_{1}(T)=\mu_{2}(T)$. Here rays III and V are not possible, unless additional condition of $\mu_1 = \mu_2$ is  fulfiled. That however leads to the existence of two degenerate minima for $T=0$.

If $\tilde c \not = 1$ then after EWSB only one $EW\! v$ minimum is created (either $I_1$ or $I_2$). However, there is a possibility that $\tilde c = 1$ and $\mu_1 = \mu_2$. In this case after EWSB two degenerate minima appear and they have the same energy for the whole time after symmetry breaking. Note, that because of the form of Yukawa interaction the fermionic contribution appears only in $c_1$ and so $\tilde c = 1$ is possible only if $\lambda_1 \not = \lambda_2$.

\begin{center}
\begin{table}[h]
\centering
{\renewcommand{\arraystretch}{1.5}

\begin{tabular}{|c|c|c|c|}
\hline \hline
Ray  & Sequence & Conditions & Co-existing min. and extr. for $T=0$\\ \hline \hline
Ia & $EW \! s \to I_1$ & $\mu_2<0$ & - \\ \hline
IIa & $EW\! s \to I_1$ & $0 < \mu_2 < \textrm{Min}\left( \mu_1 \tilde c , \mu_1 R^{-1} \right)$ & $I_2$ extremum \\ \hline
III & $EW\! s \to I_1$ & $\mu_1 R^{-1} < \mu_2 < \textrm{Min}\left( \mu_1 \tilde c, \mu_1 \right)$ & $I_2$ local minimum \\ \hline
IV & $EW\! s \to I_2 \to I_1$ & $\mu_1 \tilde c < \mu_2 < \mu_1 R^{-1}$ &  $I_2$ extremum \\ \hline
V & $EW\! s \to I_2 \to I_1$ & $\textrm{Max}\left( \mu_1 \tilde c, \mu_1 R^{-1} \right) < \mu_2 <\mu_1$ & $I_2$ local minimum \\ \hline \hline
\end{tabular}
}
  \caption{Possible rays for $R > 1$. \label{tabler1}}
\end{table}
\end{center}

\subsubsection{$0<R<1$}

In the $0<R<1$ case for every point in phase diagram ($\mu_1(T),\mu_2(T)$) there is only one existing minimum (and so it is a vacuum), as shown in fig.\ref{wykresb}, and all transitions are of the 2nd-order. We can reach the today's inert phase by a single phase transition or through the sequence of three 2nd-order phase transitions (table \ref{tabler2}. First type of sequence $EW\!  s \to I_1$ is realized by rays Ib and IIb, which are the analogs of rays Ia and IIb. For ray VI, which corresponds to the sequence $EW\!  s \to I_2 \to M \to I_1$, EWSB happens at temperature (\ref{TEWSBI_2}).  Then there are two more transitions in this sequence: from $I_2$ into $M$ and from $M$ into the inert vacuum $I_1$, with the last transition at temperature (\ref{tempIN}).

\begin{center}
\begin{table}[h]
\centering
{\renewcommand{\arraystretch}{1.5}

\begin{tabular}{|c|c|c|c|}
\hline \hline
Ray no. & Sequence & Conditions & Co-existing min. and extr. for $T=0$ \\ \hline \hline
Ib & $EW\! s \to I_1$ & $\mu_2 <0$ & - \\ \hline
IIb & $EW\! s \to I_1$ & $0<\mu_2<\textrm{Min}\left( \mu_1 \tilde c, \mu_1 R  \right)$ & $I_2$ extremum \\ \hline
VI & $EW\! s \to I_2 \to M \to I_1$ & $\mu_1  \tilde c <\mu_2<\mu_1 R$ & $I_2$ extremum \\ \hline
\hline
\end{tabular}
}
  \caption{Possible rays for $1>R>0$. \label{tabler2}}
\end{table}
\end{center}

\subsubsection{$-1<R<0$}
We consider the restoration of EW symmetry in the past (both $c_1,c_2$ positive) and there is only one possible ray Ic, which is similar to the rays Ia and Ib (table \ref{tabler3}, fig.\ref{wykresc}) \cite{nasza}.

\begin{center}
\begin{table}[h]
\centering
{\renewcommand{\arraystretch}{1.5}

\begin{tabular}{|c|c|c|c|}
\hline \hline
Ray no. & Sequence & Conditions & Co-existing min. and extr. for $T=0$\\ \hline \hline
Ic & $EW\! s \to I_1$ & $\mu_2 < \mu_1 R < 0$ & - \\ \hline
\hline
\end{tabular}
}
  \caption{Possible rays for $0>R>-1$. \label{tabler3}}
\end{table}
\end{center}

\section{Temperature evolution of physical parameters}

In the previous sections to describe the history of the Universe we used the parameters of the Lagrangian, namely $
\mu_1,\mu_2 \textrm{ and } R, \tilde{c}$.
Relations between them gave us information about the possible sequences of the phase transitions. In this section  we illustrate  the underlying temperature evolutions of the physical parameters (i.e. masses of the particles) for various rays, each representing the different history of the Universe. 
Here, it is useful to fix six free parameters of the model in form of four physical masses 
\begin{equation}
 M_{h_S},\quad M_{D_A}, \quad M_{D_H}, \quad M_{D^\pm}
\end{equation}
and two self-couplings: $\lambda_{345}$  and $\lambda_2$. The considered values of masses and $\lambda's$ are chosen in agreement with existing constraints both from the colliders and the DM abundance measurements, as discussed in sec.~\ref{sec:constr}. The values of $\Omega_{DM} h^2$ were calculated with the existing micrOMEGAs code. We expect the $\Omega_{DM} h^2$ to lie in the 3$\sigma$ WMAP allowed range:
\begin{equation}
0.085 < \Omega_{DM} h^2 < 0.139. \label{omega}
\end{equation} 
As we treat those results as an estimate only the fact that for some rays the calculated value is slightly outside the this range
does not exclude automatically those rays.

The following scalar mass set was used for the today's inert phase:
\be
M_{h_S} = 120 \textrm{ GeV}, \quad M_{D_H} = 45 \textrm{ GeV}, \quad M_{D_A} = 115 \textrm{ GeV}, \quad M_{D^\pm} = 125 \textrm{ GeV}, \label{massnum}
\ee
with self-couplings $\lambda_{345}$ and $\lambda_2$ different for each ray \cite{DSParis}\fn{The detailed discussion of the importance and constraints of those couplings is in progress \cite{Omega-DS}.}.

Below we show  mass evolutions arising from (\ref{Tempdep}) represented by  different rays for all three $R$ regions: rays Ia, III, IV and V for $R>1$, ray VI for $1>R>0$ and ray Ic for $R<0$ (figs.\ref{fig:ray1}-\ref{fig:ray9}).
We plot  the temperature dependent masses of the scalars for every vacua that is realized for a chosen ray. We also present the temperature dependence of the ``mass parameters'' for scalar states of local minima. In addition we present dependence on the temperature of two other physical parameters, namely the top quark mass $m_t(T)$ and $v(T)$ (proportional to the $W$ boson mass $M_W(T)$).

The initial state of the Universe is the EW symmetric phase with two massive scalar doublets $\Phi_S, \Phi_D$ and massless both fermions as well as gauge bosons. In this vacuum there are eight massive scalar states that come from the two scalar doublets, denoted on plots by $\Phi_S, \Phi_D$, with masses equal to $|m_{11}(T)|/\sqrt{2}, \; |m_{22}(T)|/\sqrt{2}$, respectively. 
After EWSB the Universe enters one of the symmetry violating vacuum with the proper particle content, with masses described by the formulas presented in the appendix \ref{sec:appendix}.

\subsection{The $R>1$ case}
\subsubsection{Ray Ia}

\begin{figure}[hb]
\vspace{-10pt}
  \centering
  \subfloat[masses of scalar states in $I_1$: $M_{D_H}$ (blue), $M_{D_A}$ (red), $M_{D^+}$ (purple), $M_{h_S}$ (green) and $EW\! s$: $\Phi_D$ (black), $\Phi_S$ (grey)]{\label{fig:ray1seq}\includegraphics[width=0.51\textwidth]{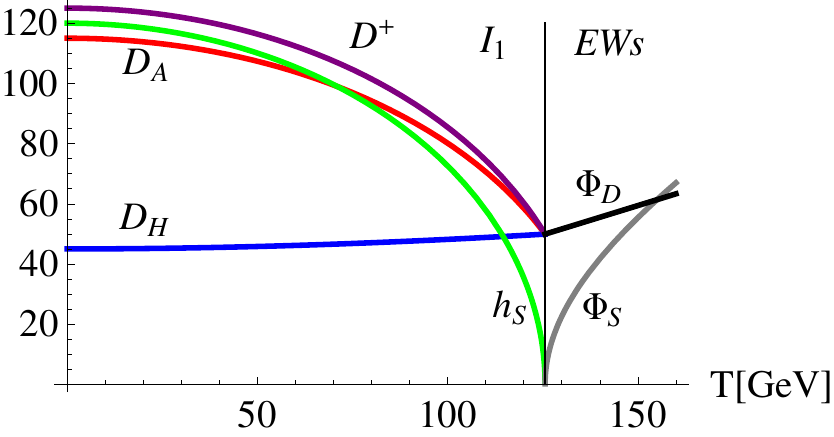}} \quad
  \subfloat[$v(T)$ (solid) and $m_t(T)$ (dashed) for $EW\! s$ (black) and $I_1$ (red)]{\label{fig:ray1mass}\includegraphics[width=0.45\textwidth]{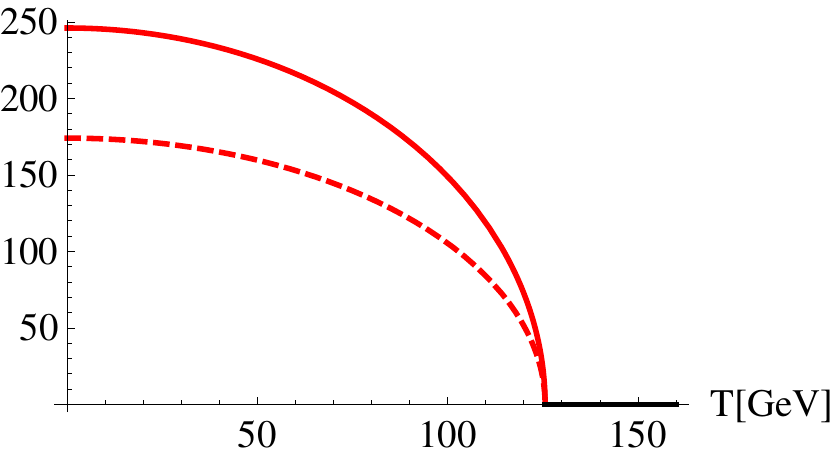}}
\vspace{-5pt}
  \caption{Evolution of masses (ray Ia): $EW\! s \to I_1$. \label{fig:ray1}}
\end{figure}

Figs.(\ref{fig:ray1}a,b) show mass evolution for the ray Ia, corresponding to the single phase transition $EW\! s \to I_1$,   for the following set of parameters:
\begin{eqnarray}
&\lambda_2 = 0.012, \quad  \lambda_{345} = 0.065 \quad  \Rightarrow \quad  c_1 = 0.913, \quad c_2 = 0.309. &
\end{eqnarray}
 The EWSB happens at $T=125.6 \textrm{ GeV}$ and Universe enters the phase with massive fermions and gauge bosons (fig.\ref{fig:ray1mass}). The  Universe stays in the $I_1$ phase with the mass of DM candidate nearly constant $M_{D_H}(T) \approx 45 \textrm{ GeV}$, fig. \ref{fig:ray1seq}.

\subsubsection{Ray III}

\begin{figure}[hb]
\vspace{-10pt}
  \centering
  \subfloat[masses of scalars in $I_1$: $M_{D_H}$ (blue), $M_{D_A}$ (red), $M_{D^+}$ (purple), $M_{h_S}$ (green) and $EW\! s$: $\Phi_D$ (black), $\Phi_S$ (grey). Dashed lines -- mass prameters of local $I_2$. Shaded region -- local minimum $I_2$.]{\label{fig:ray3seq}\includegraphics[width=0.51\textwidth]{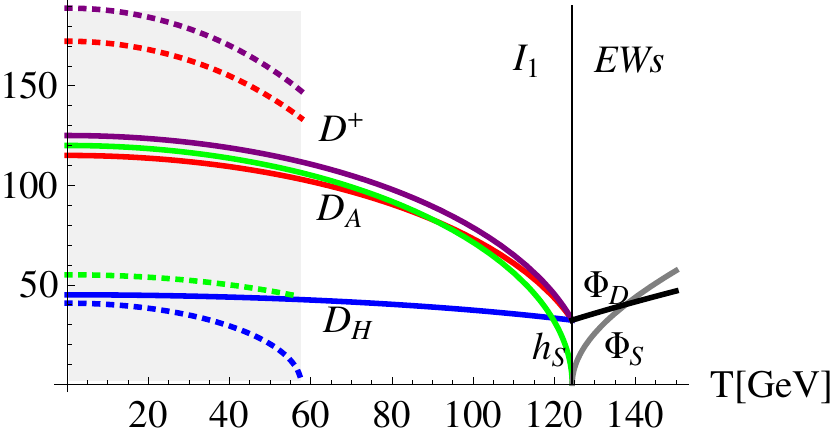}} \quad
  \subfloat[$v(T)$ (solid) and $m_t(T)$ (dashed)   for $EW\! s$ (black) and $I_1$ (red)]{\label{fig:ray3mass}\includegraphics[width=0.45\textwidth]{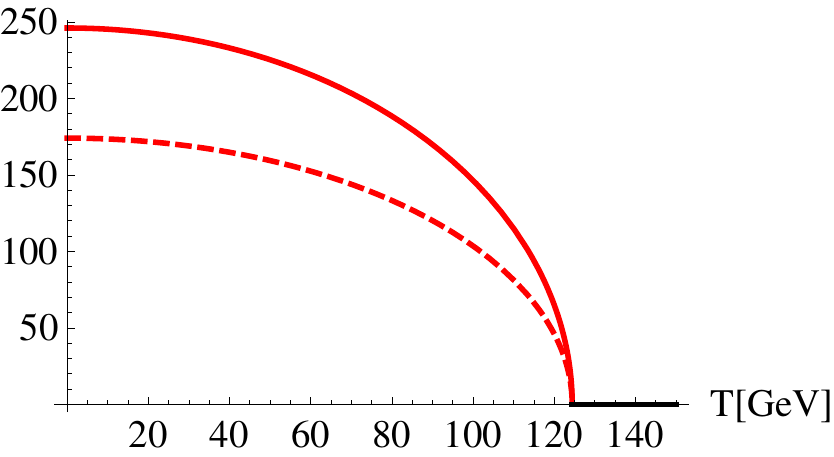}}
  \vspace{-5pt}
  \caption{Evolution of masses (ray III): $EW\!  s  \to I_1$. \label{fig:ray3}}
  \end{figure}

Ray III (figs.\ref{fig:ray3}a,b) can be realized  for:
\begin{eqnarray}
&\lambda_2 = 0.02, \quad \lambda_{345} = 0.117 \quad  \Rightarrow \quad  c_1 = 0.93, \quad c_2 = 0.33. &
\end{eqnarray}

Again here there is a single phase transition and after EWSB the Universe enters the $I_1$ phase at $T=124.5 \textrm{ GeV}$. This case is different from the previous one as at $T=57 \textrm{ GeV}$ another minimum  --  local minimum $I_2$ -- appears (shaded region in fig.\ref{fig:ray3seq}). Dashed lines show the change of "mass parameters" for corresponding scalar states of this local minimum (see appendix \ref{sec:appendix}).

\subsubsection{Ray IV}

\begin{figure}[hb]
\vspace{-10pt}
  \centering
  \subfloat[masses of scalar states around phase transitions for $I_1$ (thick lines) and $I_2$ (thin lines). Green -- $(h_S, h_D)$, purple -- $(D^\pm, S^\pm)$, red -- $(D_A,S_A)$, blue -- $(D_H,S_H)$. Dashed and dotted lines -- local minima parameters. Shaded region -- co-existence of minima $I_1$ and $I_2$.]{\label{fig:ray4seq}\includegraphics[width=0.51\textwidth]{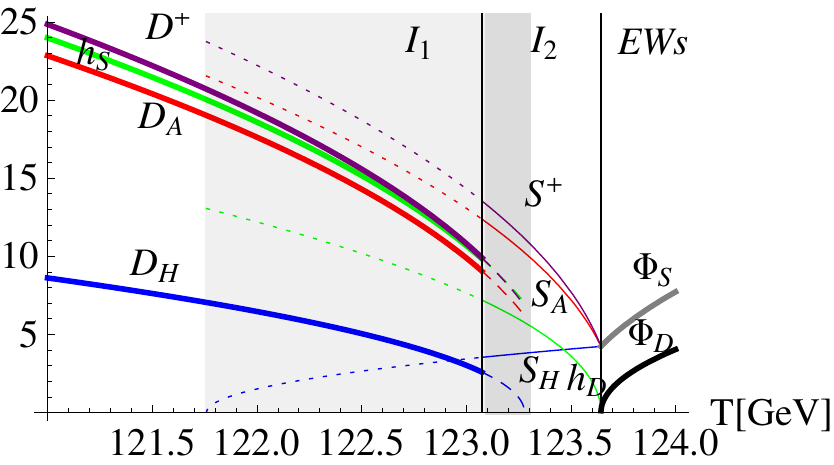}} \quad
  \subfloat[$v(T)$ (solid) and $m_t(T)$ (dashed) for $EW\! s$ (black), $I_2$ (blue) and $I_1$ (red)]{\label{fig:ray4mass}\includegraphics[width=0.45\textwidth]{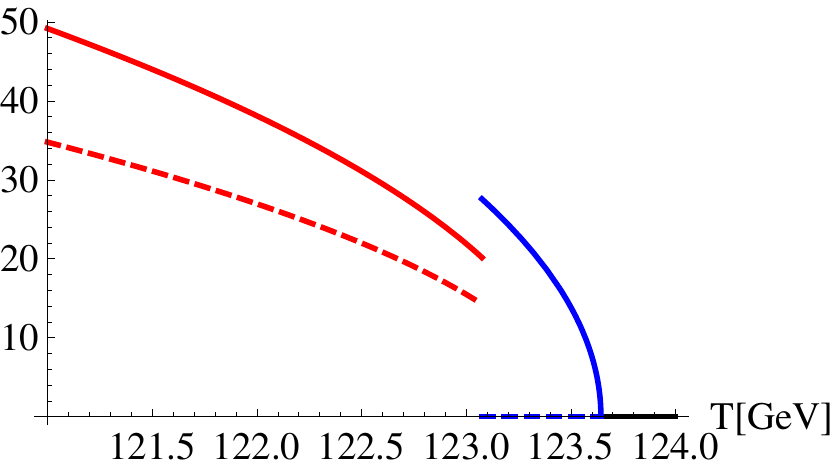}}
\vspace{-5pt}
  \caption{Evolution of masses (ray IV): $EW\! s \to I_2 \to I_1$. \label{fig:ray4}}
\end{figure}

Figs.(\ref{fig:ray4}a,b) show the mass evolution for the ray IV,  for the following set of parameters:
\begin{eqnarray}
&\lambda_2 = 0.068, \quad \lambda_{345} = 0.16 \quad \Rightarrow \quad c_1 = 0.944, \quad c_2 = 0.368. &
\end{eqnarray}

The EWSB happens at $T = 123.6 \textrm{ GeV}$ when Universe enters the inertlike phase $I_2$ with massless fermions and massive gauge bosons (fig.\ref{fig:ray4mass}). As the time grows another extremum appears, which later becomes a local minimum $I_1$. The first-order phase transition $I_2\to I_1$ happens at $T = 123.1 \textrm{ GeV}$. Note, that two minima coexist during a period of time $\Delta T \approx$ 1.5 GeV (shown by the shaded region in fig.\ref{fig:ray4seq}).  The "mass parameters" of the scalar states in the \textit{local} minima $I_1$ and $I_2$ are shown. The discontinuity in masses of physical particles: scalars, fermions and gauge boson (proportional to $v$) is visible. Universe enters the inert phase $I_1$ with massive fermions, gauge bosons and scalars, among them with DM candidate $D_H$ and their mass evolution continues up to the $T=0$ mass values.

\subsubsection{Ray V}

\begin{figure}[ht]
\vspace{-10pt}
  \centering
  \subfloat[masses of scalar states for $I_1$ (thick lines) and $I_2$ (thin lines). Green -- $(h_S, h_D)$, purple -- $(D^\pm, S^\pm)$, red -- $(D_A,S_A)$, blue -- $(D_H,S_H)$. Dashed and dotted lines -- local minima parameters. Shaded region -- co-existence of minima $I_1$ and $I_2$.]{\label{fig:ray5seq}\includegraphics[width=0.51\textwidth]{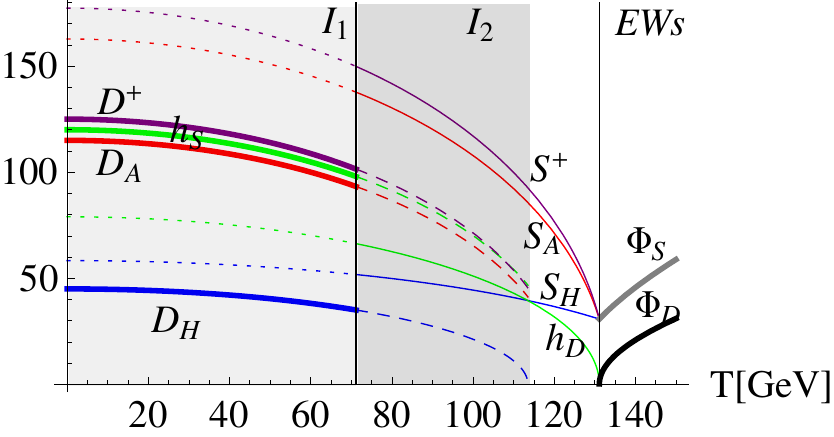}} \quad
  \subfloat[$v(T)$ (solid) and $m_t(T)$ (dashed)  for $EW\! s$ (black), $I_2$ (blue) and $I_1$ vacuum (red)]{\label{fig:ray5mass}\includegraphics[width=0.45\textwidth]{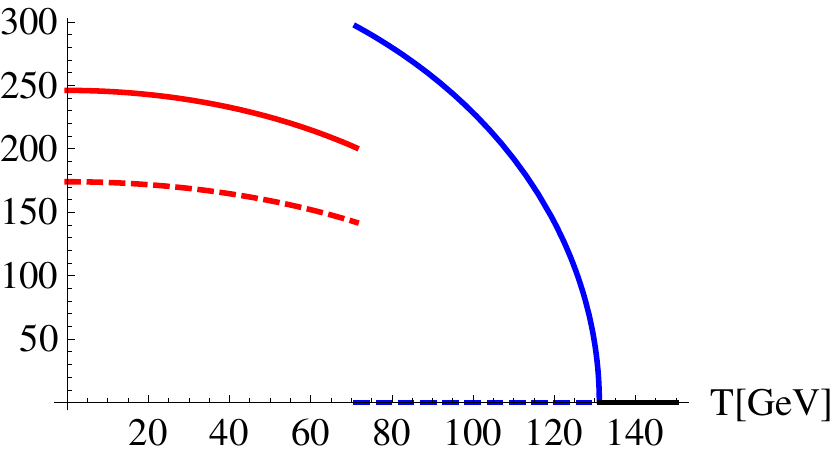}}
  \vspace{-5pt}
  \caption{Evolution of masses (ray V): $EW\! s \to I_2 \to I_1$.}
  \label{fig:ray5}
\end{figure}

Figs.(\ref{fig:ray5}a,b) show ray V, which can be realized for the following parameters:
\begin{eqnarray}
&\lambda_2 = 0.05, \quad \lambda_{345} = 0.17 \quad \Rightarrow \quad  c_1 = 0.948, \quad c_2 = 0.363. &
\end{eqnarray}

First, there is EWSB into the $I_2$ phase at $T = 131 \textrm{ GeV}$. Then at $T = 113.5 \textrm{ GeV}$ the local minimum $I_1$ appears. The first order $I_2 \to I_1$ transition happens at $T = 71 \textrm{ GeV}$, and $I_2$ becomes a local minimum, which does not disappear up to $T=0$. These two minima coexist during a period represented by the shaded region in fig.\ref{fig:ray5seq}. Dashed lines correspond to the "mass parameters" of the scalar states in the \textit{local} minima $I_1$ and $I_2$. Note, that for this ray the final phase transition happens at the lower temperatures than in the other cases, where  ratio $M_{D_A,D^\pm}/T$ is of the order 1.

\subsection{The $1>R>0$ case}

\subsubsection{Ray VI}

\begin{figure}[hb]
\vspace{-10pt}
  \centering
  \subfloat[masses of scalar states around phase transitions for $I_1$ (thick lines), $M$ (thin) and $I_2$ (dashed). Green -- $(h_S, H, h_D)$, purple -- $(D^\pm, H^\pm, S^\pm)$, red -- $(D_A,A,S_A)$, blue -- $(D_H,h,S_H)$]{\label{fig:ray8seq}\includegraphics[width=0.51\textwidth]{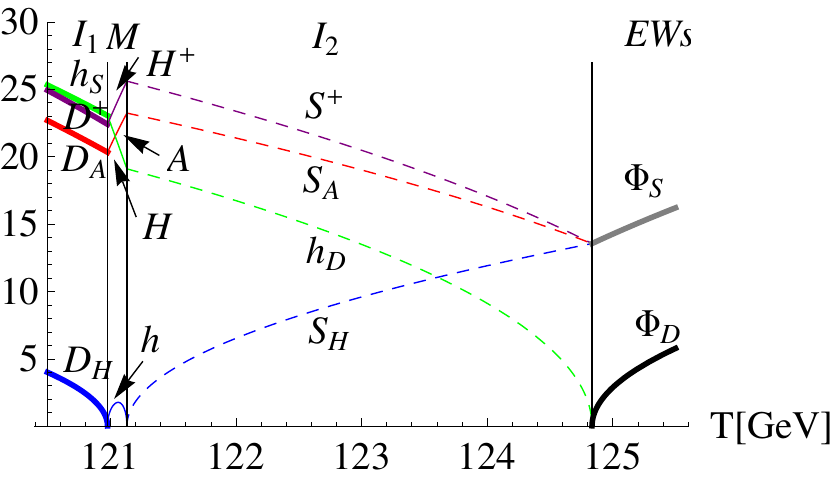}} \quad
  \subfloat[$v(T)$ (solid) and $m_t(T)$ (dashed) for $EW\! s$ (black), $I_2$ (blue), $M$ (green) and $I_1$ (red)]{\label{fig:ray8mass}\includegraphics[width=0.45\textwidth]{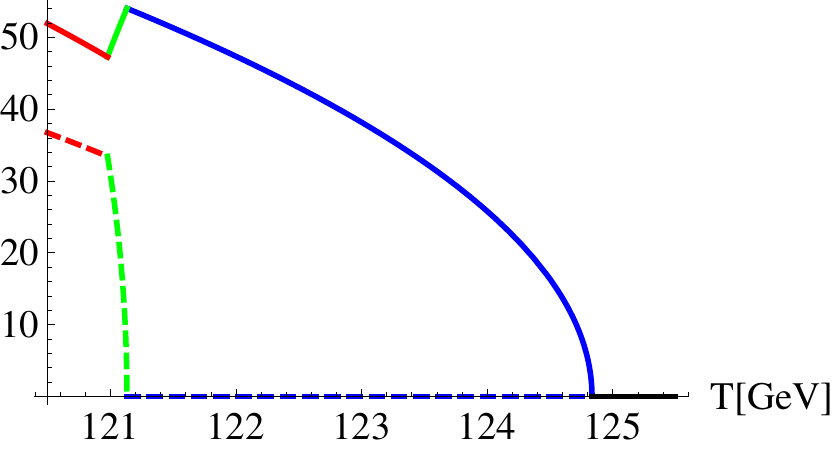}}
\vspace{-5pt}
  \caption{Evolution of masses (ray VI): $EW\!  s \to I_2 \to M \to I_1$. \label{fig:ray8}}

\end{figure}

Figs.(\ref{fig:ray8}a,b) show the evolution of the Universe along the ray VI. At every temperature there is only one minimum, all three transitions are of the 2nd-order. We study mass evolutions for:
\begin{eqnarray}
&\lambda_2 = 0.125, \quad \lambda_{345} = 0.17 \quad \Rightarrow \quad c_1 = 0.947, \quad c_2 = 0.40. &
\end{eqnarray}

EWSB happens at $T = 124.8 \textrm{ GeV}$. Universe enters the inertlike phase $I_2$  with massless fermions and massive gauge bosons (fig.\ref{fig:ray8mass}). At $T = 121.1 \textrm{ GeV}$ the mass of $S_H$ particle goes to 0 and the 2nd-order transition to the $M$ phase takes place. This phase is very short lived, at the beginning and at the end the mass of $h$ particle goes to zero, at $T = 121.1 \textrm{ GeV}$ and $T = 120.9 \textrm{ GeV}$, respectively. At this last 2nd-order transition Universe enters the inert phase $I_1$ with the DM candidate $D_H$. The mass evolution continues to the today's values of masses. Note, that here rays V and VI have the same $\lambda_{345}$, while they differ by the value of $\lambda_2$.

\subsection{The case $0>R>-1$}

\subsubsection{Ray Ic}

\begin{figure}[ht]
\vspace{-10pt}
  \centering
  \subfloat[masses of scalar states in $I_1$: $M_{D_H}$ (blue), $M_{D_A}$ (red), $M_{D^+}$ (purple), $M_{h_S}$ (green) and $EW\! s$: $\Phi_D$ (black), $\Phi_S$ (grey)]{\label{fig:ray9seq}\includegraphics[width=0.51\textwidth]{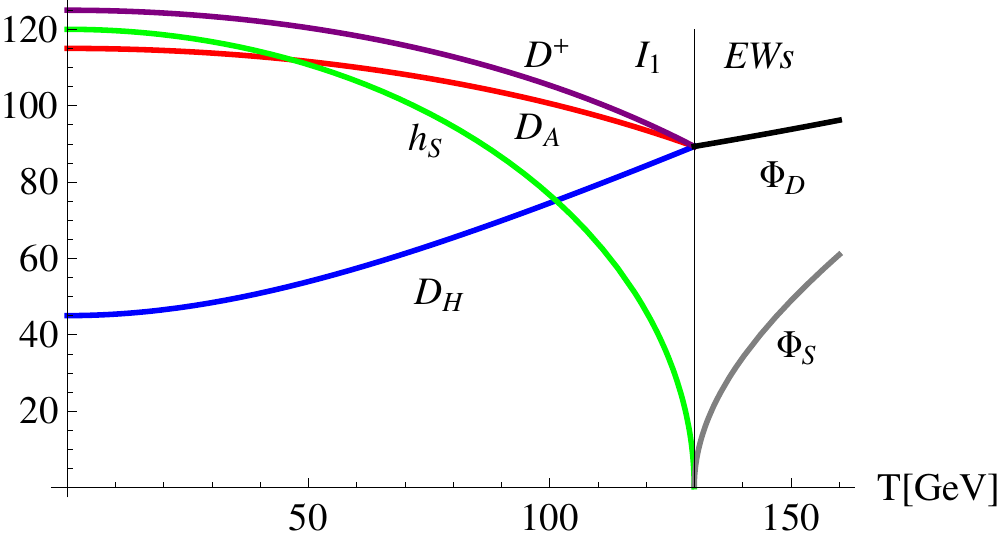}} \quad
  \subfloat[$v(T)$ (solid) and $m_t(T)$ (dashed)  for $EW\! s$ (black) and $I_1$ vacuum (red)]{\label{fig:ray9mass}\includegraphics[width=0.45\textwidth]{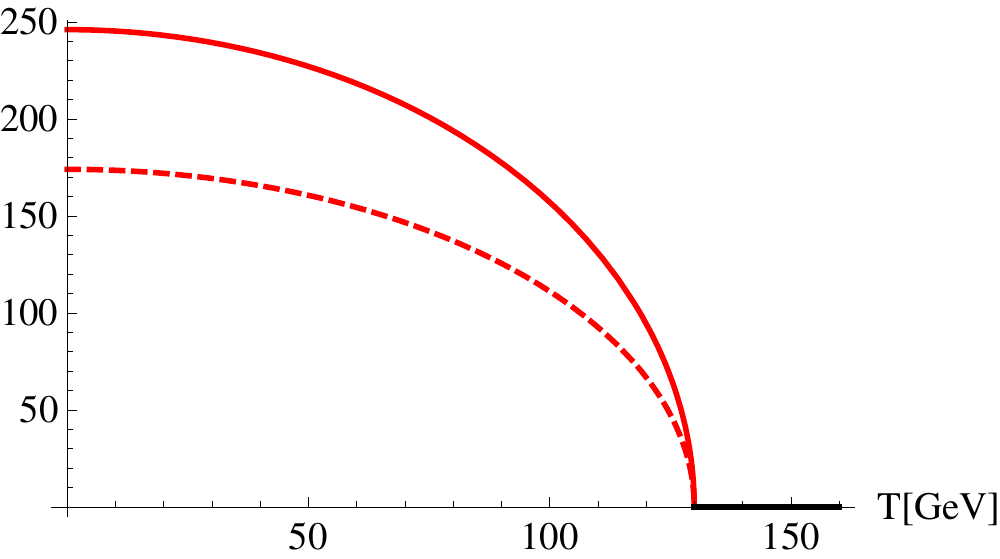}}
  \vspace{-5pt}
  \caption{Evolution of masses (ray Ic): $EW\! s \to I_1$.}
  \label{fig:ray9}
\end{figure}

Ray Ic shown in figs.(\ref{fig:ray9}a,b) is realized for:
\begin{eqnarray}
&\lambda_2 = 0.1, \quad \lambda_{345} = -0.115 \quad  \Rightarrow \quad c_1 = 0.852, \quad c_2 = 0.293. &
\end{eqnarray}

In this case there is a single phase transition (EWSB)  at $T = 130 \textrm{ GeV}$. Universe enters the inert phase $I_1$ with massive fermions, gauge bosons and DM candidate $D_H$.

\subsection{Relict densities for considered rays}

\begin{figure}[htb]
\centering
\includegraphics{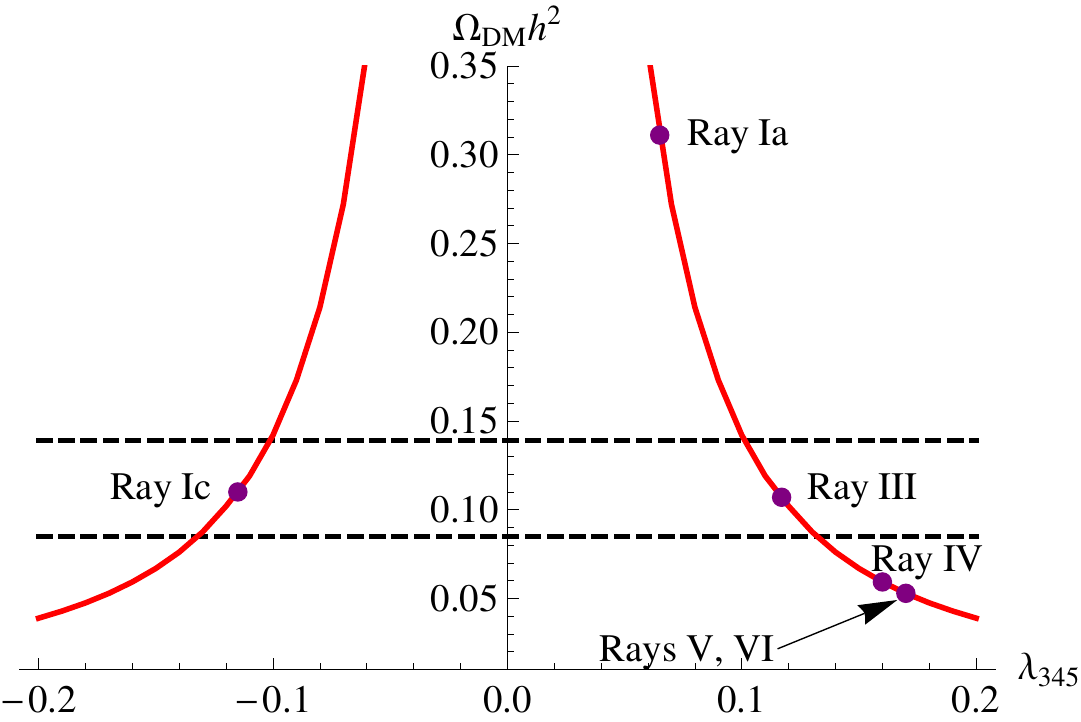}
  \caption{Relict density $\Omega_{DM} h^2(\lambda_{345})$ (red curve) with WMAP limits (dashed lines). Dots correspond to  particular rays.}
  \label{omega_45}
\end{figure}

We use the micrOMEGAs code \cite{micro} for a rough estimation of the relict density for the rays described in the previous sections 4.1-4.3. Results for $\Omega_{DM} h^2$ are presented  in fig.\ref{omega_45}, where dots represent values obtained for the different rays. Ray Ia with $\Omega_{DM} h^2 = 0.31$ is excluded by the WMAP results. For rays Ic and III values of $\Omega_{DM} h^2$ are within the WMAP range. As $|\lambda_{345}|$ grows the observed $\Omega_{DM} h^2$ decreases and for rays IV, V and VI it is below the lower WMAP limit. $\Omega_{DM} h^2$ for rays VI and V are equal,  as those rays differ only by the value of $\lambda_2$, which does not enter explicitly the rates for processes relevant for the DM abundance.

As discussed before, those values are an estimate of the real relict density. Sizable corrections are expected  especially for ray V, because in this case the temperature of the final phase transition (1st-order) is the lowest.

Fig.\ref{omega_45} contains $\Omega_{DM} h^2$ as a function of $\lambda_{345}$. Calculation was done for $\lambda_2 = 0.15$. Although $\Omega_{DM}$ does not depend explicitly of the value of $\lambda_2$,  its value limits the range of $\lambda_{345}$ we can scan over because of
the positivity constraints and  necessary conditions for the inert minimum to be a global minimum \cite{Omega-DS}. In the considered case we can have the physical solutions only in range $0.2 > \lambda_{345} > - 0.2$.

 For the large mass splitting that we chose, the coannihilation is not important and the main process is $D_H D_H \to b \bar b$. For low values of $|\lambda_{345}|$ the relict density is high, as the annihilation via $h_S$ is low. It becomes more important and it lowers the resulting $\Omega_{DM} h^2$ as $|\lambda_{345}|$ grows.
 We find that two $\lambda_{345}$ regions, $ \lambda_{345} \in (-0.105,-0.13), (0.105,0.13)$, are in agreement with the WMAP limit (\ref{omega}).

\section{Discussion and conclussions}

In this paper we studied further the evolution of the Universe after inflation towards the present inert phase. We considered different types of evolution, which may be parametrized by using the parameter $R$. 

We extend the studies with respect to the previous paper \cite{nasza} by discussing the possibility of co-existence of the local minima. This oportunity may be realized for two types of rays - ray IV and V. In the first case the co-existence is only temporary, as the phase transitions happen in a short period of time and the local minimum $I_2$ dissapears shortly after the Universe enters the inert phase. 

However, we find that also other situation is possible. For ray V the local minimum $I_2$ does not disappear and it exists for $T=0$. Also for ray III, which corresponds to a single phase transition, the local minimum $I_2$ appears for the later stages of evolution and it exists for $T=0$. Furthermore, this ray gives a good relict density.

We also stress the fact that the intermediate phases $I_2,M$ for rays IV and VI are short-lived. In this sense those sequences may be considered as a similar to the one with the single phase transition, however the latent heat of the 1st-order phase transition may affect the evolution.

One should keep in mind the limitiations of using micrOMEGAs in the context of our analysis. First, we consider only the two body final states and so, for our mass range, the actual $\Omega_{DM}$ may be lower \cite{Honorez3}. Furthermore, the temperature dependence of masses and the latent heat from the first order phase transition will influence the value of $\Omega_{DM} h^2$. We expect that the relict density for rays IV-VI will be higher, as the final phase transition happens at the lower temperature than the EWSB for the other rays. The corrections from the 1st-order phase transition will result in the splitting between rays V and VI, which for now have the same relict density. We should also take into account the fact, that during the evolution the relation between the mass of the DM particle $M_{D_H}$ and masses of the possible decay states ($M_W, M_t, M_b$) may vary. For certain rays decay channels other than $\bar{b} b$ may play important role and affect the energy relict density (for example, $D_H D_H \to WW$ for $M_{D_H}(T)> M_{W}(T)$). The detailed analysis of those effects is in preparation \cite{pracaTEMP}.

\paragraph{Acknowledgement}
I would like to thank I. Gizburg, K. Kanishev and M. Krawczyk for cooperation and discussions. We are thankful to P. Chankowski for discussions, to A. Pukhov and A. Belyaev for useful informations about micrOMEGAs package. Work was partly supported by Polish Ministry of Science and Higher Education Grant N N202 230337.

\appendix

\section{Vacua properties \label{sec:appendix}}
The neutral solution of extremum conditions give the following values of $v_S,v_D$ parameters:
  \bea
{\pmb {EW\! s}}:&  v_D=0, \quad v_S=0,\quad\quad \quad\quad {\cal E}_{EWs}=0;&\label{Sol0bas}\\
{\pmb {I_1}}:& v_D=0,\quad v_S^2=v^2=\fr{m_{11}^2}{\lambda_1},\quad
     {\cal E}_{I_1}=-\fr{m_{11}^4}{8\lambda_1};&
     \label{solAbas}\\
{\pmb {I_2}}:& v_S=0,\quad v_D^2=v^2=\fr{m_{22}^2}{\lambda_2},\quad
     {\cal E}_{I_2}=-\fr{m_{22}^4}{8\lambda_2};&
     \label{solBbas}\\
{\pmb M}:&
    \begin{array}{c}
      v_S^2=\fr{m_{11}^2\lambda_2-\lambda_{345}m_{22}^2}{\lambda_1\lambda_2-\lambda_{345}^2},\quad
      v_D^2=\fr{m_{22}^2\lambda_1-\lambda_{345}m_{11}^2}{\lambda_1\lambda_2-\lambda_{345}^2};\\[4mm]
      {\cal E}_{M}=-\fr{m_{11}^4\lambda_2-2\lambda_{345}m_{11}^2m_{22}^2+m_{22}^4\lambda_1}
      {8(\lambda_1\lambda_2-\lambda_{345}^2)}.
    \end{array}&
\label{Nextr1}
\eea

Some of the equations \eqref{solAbas}-\eqref{Nextr1} can give also negative values
of  $v_S^2$ or $v_D^2$. In  such case the extremum, described by corresponding equations,
is absent.

\subsection{EW symmetric vacuum $EW\! s$}

The electroweak symmetric extremum $EW\! s$ (\ref{Sol0bas}) is a minimum if $m_{11,22}^2 <0$. Gauge bosons and fermions are massless, while the doublets have non-zero masses $\frac{|m_{11}^2|}{\sqrt{2}}$ and $\frac{|m_{22}^2|}{\sqrt{2}}$, respectively.

\subsection{Inert-like vacuum $I_2$}\label{secinertlike}

\emph{The inertlike vacuum} $I_2$  is "mirror-symmetric" to the inert vacuum
$I_1$, compare \eqref{solAbas} and \eqref{solBbas}, with  one Higgs particle $h_D$ and  four scalar particles: $S_H,\,S_A,\,S^\pm$. The interaction among scalars and between scalars and gauge bosons  are
mirror-symmetric as well, so the only
difference between $I_2$ and $I_1$ arises from the Yukawa interaction.

The inertlike vacuum $I_2$  violates
 $D$-symmetry. The Higgs boson $h$ couples to gauge bosons just as the Higgs boson of the SM, however
it does not couple to fermions at the tree level. The scalars do interact with fermions. Therefore, here there are  no candidates for dark matter particles. Note that all fermions, by definition interacting only with $\Phi_S$ with vanishing v.e.v.
 $\la\Phi_S\ra =0$,  are massless. (Small mass can appear only as a loop effect.)

The masses of the Higgs boson $h_D$  and $S$-scalars
are given by (cf. \eqref{massesA}) with $m_{11}^2 \leftrightarrow m_{22}^2$.


\subsection{Mixed vacuum $M$}

The mixed extremum\fn{Sometimes called  a\emph{ normal extremum} $N$, see e.g. \cite{lorenzo}}  $M$  violates the $Z_2$ symmetry. In this vacuum we have massive fermions and no candidates for DM particle, like in the SM. There are five Higgs bosons - two charged $H^\pm$ and three neutral ones: the CP-even $h$ and $H$ and CP-odd  $A$. Couplings of the physical Higgs bosons to fermions and gauge bosons have
standard forms as for the  2HDM, with the Model I Yukawa interaction.

%

 Masses of scalars are as follows (see, e.g.
  \cite{GK05,GK07})

\be
 M_{H^\pm}^2=-\fr{\lambda_4+\lambda_5}{2}v^2\,,\quad
 M_A^2=-v^2\lambda_5,\quad \left(v^2=v_S^2+v_D^2\right). \label{chneitr}
\ee
The neutral CP-even  mass matrix is equal to
\be
 {\cal{M}}=\begin{pmatrix}\lambda_1v_S^2&\lambda_{345}v_Sv_D\\
                             \lambda_{345}v_Sv_D&\lambda_2v_D^2\end{pmatrix}\,.
\label{massmatrixCV}
\ee

The  mass matrix \eqref{massmatrixCV} gives masses of the neutral CP-even Higgs bosons:
\be
 M_{h,H}^2=\fr{\lambda_1v_S^2+\lambda_2v_D^2\pm
 \sqrt{(\lambda_1v_S^2+\lambda_2v_D^2)^2-4\det{\cal M}}}{2}\,,\label{massesC}
\ee
with sign $+$ for the $H$ and sign $-$ for $h$.


\begin{thebibliography}{99}

\bibitem{inert}
  N.~G.~Deshpande and E.~Ma,
  Phys.\ Rev.\  D {\bf 18} (1978) 2574;
  R.~Barbieri, L.~J.~Hall and V.~S.~Rychkov,
  Phys.\ Rev.\  D {\bf 74} (2006) 015007
  [arXiv:hep-ph/0603188].

\bibitem{GK07}
  I.~F.~Ginzburg and K.~A.~Kanishev,
  Phys.\ Rev.\  D {\bf 76} (2007) 095013
  [arXiv:0704.3664 [hep-ph]].
  
  
\bibitem{GIK09}
  I.~F.~Ginzburg, I.~P.~Ivanov and K.~A.~Kanishev,
  Phys.\ Rev.\  D {\bf 81} (2010) 085031
  [arXiv:0911.2383 [hep-ph]].

\bibitem{iv2008}
  I.~P.~Ivanov,
  Acta Phys.\ Polon.\  B {\bf 40} (2009) 2789
  [arXiv:0812.4984 [hep-ph]].

\bibitem{nasza}
I. F. Ginzburg, K.A. Kanishev, M. Krawczyk, D. Sokolowska, Phys. Rev. D 82, 123533 (2010)
[arXiv:1009.4593 [hep-ph]]

\bibitem{Omega-DS} D. Soko\l owska: Dark Matter data and constraints on the quartic coupling (work in progress)

\bibitem{micro}
G. Belanger, F. Boudjema, A. Pukhov and A. Semenov, Comput. Phys. Commun. 176
(2007) 367 [arXiv:hep-ph/0607059]. G. Belanger, F. Boudjema, A. Pukhov and A. Semenov, arXiv:0803.2360 [hep-ph]. G. Belanger, F. Boudjema, A. Pukhov and A. Semenov, Comput. Phys. Commun. 174 (2006) 577 [arXiv:hep-ph/0405253]. G. Belanger,
F. Boudjema, A. Pukhov and A. Semenov, Comput. Phys. Commun. 149 (2002) 103
[arXiv:hep-ph/0112278].



\bibitem{limpap}
  Q.~H.~Cao, E.~Ma and G.~Rajasekaran,
  Phys.\ Rev.\  D {\bf 76} (2007) 095011
  [arXiv:0708.2939 [hep-ph]];
  P.~Agrawal, E.~M.~Dolle and C.~A.~Krenke,
  Phys.\ Rev.\  D {\bf 79}, 015015 (2009)
  [arXiv:0811.1798 [hep-ph]];
  M.~Gustafsson, E.~Lundstrom, L.~Bergstrom and J.~Edsjo,
  Phys.\ Rev.\ Lett.\  {\bf 99} (2007) 041301
  [arXiv:astro-ph/0703512];


\bibitem{Dolle:2009fn}
  E.~M.~Dolle and S.~Su,
  Phys.\ Rev.\  D {\bf 80} (2009) 055012
  [arXiv:0906.1609 [hep-ph]];
  E.~Dolle, X.~Miao, S.~Su and B.~Thomas,
  Phys.\ Rev.\  D {\bf 81}, 035003 (2010)
  [arXiv:0909.3094 [hep-ph]];

\bibitem{LopezHonorez:2006gr}
  L.~Lopez Honorez, E.~Nezri, J.~F.~Oliver and M.~H.~G.~Tytgat,
  JCAP {\bf 0702} (2007) 028
  [arXiv:hep-ph/0612275]; 
  C.~Arina, F.~S.~Ling and M.~H.~G.~Tytgat,
  JCAP {\bf 0910} (2009) 018
  [arXiv:0907.0430 [hep-ph]];
  T.~Hambye and M.~H.~G.~Tytgat,
  Phys.\ Lett.\  B {\bf 659} (2008) 651
  [arXiv:0707.0633 [hep-ph]];
  E.~Nezri, M.~H.~G.~Tytgat and G.~Vertongen,
  JCAP {\bf 0904} (2009) 014
  [arXiv:0901.2556 [hep-ph]];
  S.~Andreas, M.~H.~G.~Tytgat and Q.~Swillens,
  JCAP {\bf 0904} (2009) 004
  [arXiv:0901.1750 [hep-ph]];
  S.~Andreas, T.~Hambye and M.~H.~G.~Tytgat,
  JCAP {\bf 0810} (2008) 034
  [arXiv:0808.0255 [hep-ph]];
  L.~L.~Honorez and C.~E.~Yaguna,
  arXiv:1003.3125 [hep-ph];

\bibitem{Lundstrom:2008ai}
  E.~Lundstrom, M.~Gustafsson and J.~Edsjo,
  Phys.\ Rev.\  D {\bf 79} (2009) 035013
  [arXiv:0810.3924 [hep-ph]].
  
  
\bibitem{kra-sok}
  M.~Krawczyk and D.~Soko\l owska,
  arXiv:0911.2457 [hep-ph].

\bibitem{PDG} Particle Data Group. {\it Journ. of Phys.} {\bf G 37} \#7A (2010) 075021

\bibitem{pracaTEMP} I. Ginzburg (private communication); I.Ginzburg, K.Kanishev, M.Krawczyk, D.Soko\l owska (work in progress)

\bibitem{Gavela:1998ux}
  M.~B.~Gavela, O.~Pene, N.~Rius and S.~Vargas-Castrillon,
  Phys.\ Rev.\  D {\bf 59} (1999) 025008
  [arXiv:hep-ph/9801244];
%

\bibitem{lorenzo}
  J.~L.~Diaz-Cruz and A.~Mendez,
  Nucl.\ Phys.\  B {\bf 380} (1992) 39.
  P.~M.~Ferreira, R.~Santos and A.~Barroso,
  Phys.\ Lett.\  B {\bf 603}, 219 (2004)
  [Erratum-ibid.\  B {\bf 629}, 114 (2005)]
  [arXiv:hep-ph/0406231];
  A.~Barroso, P.~M.~Ferreira and R.~Santos,
  Phys.\ Lett.\  B {\bf 632}, 684 (2006)
  [arXiv:hep-ph/0507224].
  A.~Barroso, P.~M.~Ferreira and R.~Santos,
  Phys.\ Lett.\  B {\bf 652}, 181 (2007)
  [arXiv:hep-ph/0702098];
  A.~Barroso, P.~M.~Ferreira, R.~Santos and J.~P.~Silva,
  Phys.\ Rev.\  D {\bf 74} (2006) 085016
  [arXiv:hep-ph/0608282].

 \bibitem{GK05}
  I.~F.~Ginzburg and M.~Krawczyk,
  Phys.\ Rev.\  D {\bf 72} (2005) 115013
  [arXiv:hep-ph/0408011].


\bibitem{DSParis}
D. Soko\l owska, arXiv:1009.5099v1 [hep-ph]



\bibitem{Honorez3} L. L. Honorez, C. E. Yaguma, arXiv:1011.1411 [hep-ph]


\end{thebibliography}
\end{document}